\documentclass{article}
\usepackage{spconf,amsmath,amssymb,graphicx,hyperref}
\hypersetup{hidelinks}

\title{Interpreting hierarchical organisation of speaker embeddings}

\name{Yanze Xu$^{1}$, Wenwu Wang$^{1}$, Mark D. Plumbley$^{2}$}
\address{$^{1}$Centre for Vision, Speech and Signal Processing, University of Surrey, Guildford, UK\\
$^{2}$Department of Informatics, King's College London, London, UK}

\begin{document}

\maketitle

\begin{abstract}

Speaker recognition neural networks recognise speaker identities from input utterances by learning latent representations (i.e.\ speaker embeddings). However, these networks' internal mechanisms remain largely opaque, motivating research in explainable artificial intelligence (XAI) to understand them. Existing studies have analysed how speaker embeddings are organised, but rarely frame these analyses within XAI. This work proposes to explain and interpret the organisation of speaker embeddings from an XAI perspective.

To this end, we apply a hierarchical clustering algorithm, Single-Linkage Clustering (SLINK), to analyse whether our prepared speaker embeddings naturally form clusters with hierarchical relationships. The resulting hierarchical organisation (i.e.\ hierarchical clusters) is evaluated using the Cluster--Class Matching (CCM) method. Moreover, we propose a new method, termed Hierarchical Cluster--Class Matching (HCCM), to identify which hierarchical clusters best match individual semantic classes like \emph{male} and conjunctive semantic classes like \emph{UK \& male}, thereby interpreting the clusters using their matched classes. We quantify the matching degree with a new metric called the L-score, which makes imperfect matches diagnosable. HCCM's results reveal that the hierarchical clusters analysed by SLINK are well interpreted using classes related to speaker identity, gender, and nationality, showing the semantics inside the hierarchical organisation of the examined speaker embeddings.

\end{abstract}
\begin{keywords}
Explainable AI, Speaker Recognition, Representation Learning, Hierarchical Clustering, Cluster--Class Matching
\end{keywords}

\section{Introduction}
\label{sec:intro}
Explainable artificial intelligence (XAI) aims to make the decision-making processes of trained models, particularly neural networks, more understandable to humans~\cite{xu2026explainable1, xu2026explainable2}. In modern speaker recognition~\cite{nagrani2017voxceleb,chung2018voxceleb2}, a trained network maps an utterance input to a speaker embedding, i.e., a latent representation used by the network to distinguish speaker identities.

Existing XAI techniques are usually designed to explain what a recognition network considers important, or is sensitive to, when mapping an input to an output~\cite{xu2026explainable2}. In contrast, some studies have revealed how network representations are organised but are rarely framed within XAI~\cite{li2024efficient, rauber2016visualizing}. To address this gap, we propose to explain and interpret the organisation of speaker embeddings from an XAI perspective, specifically focusing on their hierarchical organisation.

Previous studies reveal the organisation of network representations using dimensionality reduction~\cite{mcinnes2018umap} or flat clustering methods~\cite{manning2008flat}. Dimensionality reduction methods project representations into low-dimensional spaces, where visually separable groups may emerge~\cite{li2024efficient}. Moreover, flat clustering methods~\cite{manning2008flat} analyse whether representations are naturally organised into flat, independent clusters (i.e.\ flat representation clusters)~\cite{carbonnelle2020intraclass, Peiffer2021}. However, these studies leave an open research question: whether the resulting flat representation clusters exhibit deeper relationships that are overlooked, such as hierarchical relationships.

In the present paper, we extract a set of speaker embeddings by feeding some prepared utterances into a trained speaker recognition network. We then use a hierarchical clustering method~\cite{mullner2011modern}, Single-Linkage Clustering (SLINK)~\cite{gower1969minimum}, to analyse whether these representations naturally form clusters with hierarchical relationships (i.e.\ hierarchical representation clusters). The resulting clusters collectively constitute the hierarchical organisation of the examined representations (i.e.\ hierarchical representation organisation). To evaluate the resulting clusters, we apply the existing Cluster--Class Matching (CCM) method~\cite{rosenberg2007v}, which measures the overall matching degree between the clusters and predefined semantic classes (i.e.\ classes pre-labelled for our utterances and shared by their corresponding representations).

Additionally, we propose a new method called Hierarchical Cluster--Class Matching (HCCM) to find one-to-one matches between the above hierarchical representation clusters analysed by SLINK and predefined semantic classes, thereby interpreting each cluster using its matched class. In HCCM, predefined semantic classes describing individual attributes of utterances and their corresponding representations (e.g.\ \emph{male}) are referred to as \textit{individual semantic classes} and can be combined using conjunction logic to form \textit{conjunctive semantic classes} (e.g.\ \emph{UK \& male}). The formation of conjunctive classes provides opportunities to interpret more hierarchical representation clusters. Moreover, HCCM introduces a new metric called the Liebig's score (L-score) to quantify the degree of each matched pair comprising a cluster and a semantic class (i.e.\ a cluster--class pair). The L-score determines the matching degree by the weaker of precision and recall, making it convenient to diagnose whether an imperfect match is primarily limited by precision or recall.

Finally, CCM results show that our hierarchical representation organisation analysed by SLINK achieves a good evaluation score. This organisation is visualised as a tree-like dendrogram, and HCCM results further show that hierarchical representation clusters displayed at different positions in the dendrogram are well interpreted using predefined semantic classes related to speaker identity, gender, and nationality.

\section{Related Work}
\label{sec:related}
The interpretability of network representations has been well studied in networks trained for generation tasks, where changes in latent representations are interpreted through observable semantic changes in generated outputs~\cite{shen2021closed}. However, this route is difficult to apply to speaker recognition networks because their representations are used to distinguish speaker identities rather than to generate outputs whose semantic changes can be directly observed. Instead, we adopt an approach that first analyses the organisation of speaker embeddings in terms of clusters and then interprets the representations at the cluster level.

Hierarchical clustering algorithms are commonly applied to speaker embeddings in speaker diarisation, which addresses the `who spoke when' problem~\cite{garcia2017speaker, singh2023supervised}. However, the hierarchical representation organisations produced in these diarisation works serve only as intermediate results, from which a subset of clusters without hierarchical relationships is selected for the final diarisation output. Consequently, the hierarchical representation organisations they produce are not the research focus, being rarely visualised or semantically interpreted, both of which are addressed in this XAI work.



Both CCM~\cite{rosenberg2007v} and the proposed HCCM method involve matching hierarchical clusters with predefined semantic classes but differ in two main aspects. Firstly, CCM focuses on calculating an overall matching score to evaluate the hierarchical clusters~\cite{zhao2005hierarchical}, while HCCM focuses on finding one-to-one matches with the aim of semantically interpreting more hierarchical clusters using their matched class. Secondly, CCM does not explicitly distinguish which types of predefined semantic classes are involved in the matching process~\cite{zhao2005hierarchical, rosenberg2007v}. In contrast, HCCM explicitly defines both individual and conjunctive semantic classes, with the latter constructed using conjunction logic to interpret more hierarchical representation clusters.

\section{Methodology}
\label{sec:method}
\subsection{Analysing and Evaluating Hierarchical Representation Organisation}
\label{sec:hierarchical_analysis}

Let $f(\cdot)$ denote a trained speaker recognition network, and let $X=\{x_i\}$ denote a set of utterances drawn from a dataset disjoint from the training data of $f$. Their corresponding representations (i.e.\ speaker embeddings) are denoted by $A=\{a_i\}$, where $a_i=f(x_i)$.

We apply one of the most popular hierarchical clustering algorithms, Single-Linkage Clustering (SLINK)~\cite{gower1969minimum}, to analyse whether the representations in $A$ exhibit a hierarchical representation organisation, in which they naturally form clusters with hierarchical relationships. The resulting clusters are collectively denoted by $\mathcal{H}=\{h_i\}$, where each $h_i$ denotes the $i$-th hierarchical representation cluster and contains the indices of all representations belonging to this cluster. SLINK is characterised by defining the distance between two representation clusters $h_p$ and $h_q$ as that between the closest pair of representations from two clusters, as follows:
\[
\text{dist}(h_p,h_q)
=
\min_{\substack{i\in h_p\\j\in h_q}}
\text{dist}(a_i,a_j),
\]
\noindent where $\text{dist}(\cdot,\cdot)$ denotes the distance between two representations; we use Euclidean distance in this work. To evaluate the hierarchical representation organisation $\mathcal{H}$, we apply the Cluster--Class Matching (CCM) method~\cite{rosenberg2007v} to measure its overall matching performance against a set of predefined semantic classes $C=\{c_i\}$. Each $c_i$ denotes the $i$-th predefined semantic class and contains the indices of all representations in $A$ whose corresponding utterances are pre-labelled with this class. For a predefined semantic class $c \in C$ and a hierarchical representation cluster $h \in \mathcal{H}$, the precision and recall of this cluster-class pair are defined as follows:
\[
P(c,h)
=
\frac{|c\cap h|}{|h|},
\qquad
R(c,h)
=
\frac{|c\cap h|}{|c|},
\]
and their F-score matching degree is quantified as follows:
\[
F(c,h)
=
\frac{2P(c,h)R(c,h)}
{P(c,h)+R(c,h)}
=
\frac{2|c\cap h|}
{|c|+|h|}.
\]
On this basis, the best-matched hierarchical representation cluster of an predefined semantic class is the one achieving the highest F-score among all clusters in $\mathcal{H}$. CCM lastly aggregates the F-scores of all best-matched cluster--class pairs to measure the overall matching performance between $C$ and $\mathcal{H}$, defined as follows:
\begin{equation}
\mathrm{CCM}(C,\mathcal{H})
=
\sum_{c\in C}
\frac{|c|}{|A|}
\max_{h\in\mathcal{H}}
F(c,h).
\label{eq:ccm}
\end{equation}
It is assumed that a higher CCM overall matching degree corresponds to a better hierarchical representation organisation $\mathcal{H}$, as hierarchical representation clusters align better with the predefined class groupings of the representations.

\subsection{Semantically Interpreting Hierarchical Representation Organisation}
\label{sec:hccm}
Instead of evaluating the above hierarchical representation clusters $\mathcal{H}$ using a single overall matching degree as in CCM, the proposed HCCM method semantically interprets the clusters in $\mathcal{H}$ by finding their one-to-one matches with predefined semantic classes.

HCCM first defines \textit{individual semantic classes} (i.e.\ \textit{individual classes}) as pre-labelled semantic classes describing a single attribute of model inputs or their representations. For example, \emph{male} is an individual class related to gender (i.e.\ a gender-related individual class), while \emph{UK} is an individual class related to nationality (i.e.\ a nationality-related individual class). We denote by $C_{\mathrm{ind}}={c_i}$ the collection of all individual classes pre-labelled for the representations $A$, where each $c_i$ denotes the $i$-th individual class and contains the indices of all representations in $A$ whose corresponding model inputs belong to this class.


HCCM next defines a \textit{conjunctive semantic class} (i.e.\ \textit{conjunctive class}) as a class formed by combining individual classes using conjunction logic, such that a model input or its corresponding representation belongs to the conjunctive class if and only if it belongs to all of its constituent individual classes. Accordingly, the index set of a conjunctive class is obtained by intersecting the index sets of its constituent individual classes. For example, the conjunctive class \emph{UK \& male} is formed as
$c_{\mathrm{UK\&male}} = c_{\mathrm{UK}} \cap c_{\mathrm{male}}$. We denote by $C_{\mathrm{conj}}$ the collection of conjunctive classes constructed by HCCM from $C_{\mathrm{ind}}$.

To provide HCCM with opportunities to match more clusters using both individual classes and the newly constructed conjunctive classes, we combine $C_{\mathrm{ind}}$ and $C_{\mathrm{conj}}$ into a larger set, denoted by $C^{(0)} = C_{\mathrm{ind}} \cup C_{\mathrm{conj}}$. HCCM then iteratively finds cluster--class pairs between $C^{(0)}$ and $\mathcal{H}$ as follows:

\begin{equation}
\begin{aligned}
(c_l^{*},h_l^{*})
&=
\arg\max_{\substack{
c\in C^{(l-1)}\\
h\in\mathcal{H}}}
L(c,h),\\
C^{(l)}
&=
C^{(l-1)}
\setminus
\{c_l^{*}\},
\qquad
l=1,\ldots,|C^{(0)}|.
\end{aligned}
\label{eq:hccm}
\end{equation}
At the $l$-th iteration, HCCM selects the globally best-matched cluster--class pair $(c_l^{*},h_l^{*})$ among all possible pairs formed by the remaining semantic classes in $C^{(l-1)}$ and the full set of hierarchical representation clusters in $\mathcal{H}$. The matched semantic class $c_l^{*}$ is then removed from $C^{(l-1)}$ before the next iteration. After $|C^{(0)}|$ iterations, HCCM produces $|C^{(0)}|$ cluster--class pairs, which semantically interpret up to $|C^{(0)}|$ distinct clusters. The L-score $L(c,h)$ in Eq.~\ref{eq:hccm} is defined as follows:
\begin{equation}
L(c,h)
=
\min\bigl(P(c,h),R(c,h)\bigr)
=
\frac{|c\cap h|}
{\max(|c|,|h|)}.
\label{eq:lscore}
\end{equation}
The design of the L-score is inspired by Liebig's law of the minimum~\cite{de1994liebig}, which suggests that system performance is constrained by its most limiting component. Accordingly, the L-score assumes that the matching performance of a cluster--class pair is limited by the weaker of precision and recall, making imperfect matches easier to diagnose. As an example, if $P(c,h)=0.9$ and $R(c,h)=0.6$, the L-score is 0.6 and is therefore limited by recall. This 0.60 indicates that 40\% of the representations belonging to semantic class $c$ are not retrieved by the matched cluster $h$. In contrast, the F-score is 0.72, which can only be interpreted as the harmonic mean of precision and recall, while what 0.72 itself represents remains difficult to interpret, as also noted by Christen et al.~\cite{christen2023review}.


\section{Experiments}
\label{sec:experiments}

\subsection{Experimental Setup}
\label{sec:experimental_setup}

The speaker recognition network used in our experiments is the ResNetSE34L model implemented by Chung et al.~\cite{chung2020defence}, trained using the angular prototypical loss~\cite{chung2020defence} on spectrograms calculated from utterances in the VoxCeleb2 development set~\cite{chung2018voxceleb2}. We input $4$-second utterances from the VoxCeleb1 test set~\cite{nagrani2017voxceleb}, corresponding to 400 $\times$ 10\,ms spectrogram frames, into the trained network and use the outputs of its penultimate layer as speaker embeddings. SLINK is then applied to these speaker embeddings to analyse their hierarchical representation organisation. Such organisation is evaluated by CCM and interpreted by HCCM. The VoxCeleb1 test set contains utterances pre-labelled with 40 speaker identities, 2 gender classes, and 9 nationality classes, yielding 51 predefined individual semantic classes. 13 conjunctive classes are constructed from the non-empty pairwise intersections of the gender-related and nationality-related individual classes (i.e.\ 5 of the $2 \times 9$ conjunctive classes are empty). Hence, 64 predefined semantic classes are available for CCM and HCCM.


\subsection{Evaluation Results of Hierarchical Representation Organisation}
\label{sec:ccm_results}

The CCM method based on the F-score metric, as discussed in Section~\ref{sec:hierarchical_analysis}, evaluates the entire hierarchical representation organisation (i.e.\ hierarchical representation clusters) analysed by SLINK, with the results reported in Table~\ref{tab:ccm_results}. Separate CCM scores are reported for identity-, gender-, and nationality-related individual classes, as well as HCCM-constructed nationality\&gender conjunctive classes, with each score calculated exclusively using the corresponding type of predefined semantic class.

\begin{table}[t]
\caption{CCM evaluation of our SLINK-analysed hierarchical representation organisation.}
\label{tab:ccm_results}
\centering
\small
\begin{tabular}{lcc}
\hline
Semantic classes & \# Classes & CCM F-score \\
\hline
Speaker identity & 40 & 0.9754 \\
Gender & 2 & 0.9997 \\
Nationality & 9 & 0.7710 \\
Nationality\&gender conjunction & 13 & 0.8316 \\
\hline
Average & -- & \textbf{0.8944} \\
\hline
\end{tabular}
\end{table}

As shown in Table~\ref{tab:ccm_results}, the hierarchical representation organisation analysed by SLINK achieves an average CCM score of 0.8944 across the four types of predefined semantic classes. In particular, identity-related individual classes achieve a very high CCM score of 0.9754. This strong alignment between identity-related individual classes (i.e.\ predefined groupings of representations into identity-related individual classes) with the resulting representation organisation is expected, since our speaker recognition network is trained to distinguish speaker identities. Moreover, gender- and nationality-related individual classes separately achieve high CCM scores of 0.9997 and 0.7710, indicating that the representation organisation also aligns very closely with these classes, even though the examined speaker recognition network is not trained for either gender or nationality recognition. Lastly, nationality\&gender conjunctive classes achieve a higher CCM score of 0.8316 than the 0.7710 achieved by nationality-related individual classes, suggesting that these conjunctive classes are more suitable for evaluating the SLINK-analysed representation organisation.


\subsection{Interpretation Results of Hierarchical Representation Organisation}
\label{sec:hccm_results}
\begin{figure}[t]
    \centering
    \includegraphics[width=\linewidth]{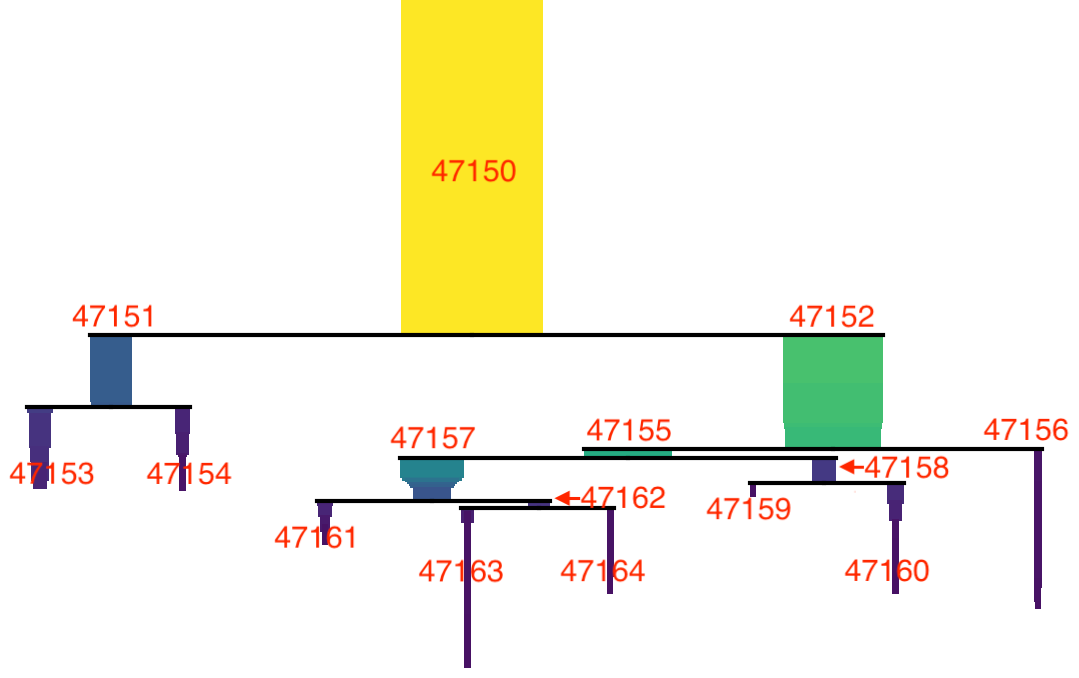}
    \caption{An icicle dendrogram visualising our SLINK-analysed hierarchical organisation of speaker embeddings. Hierarchical clusters are labelled by their cluster IDs.}
    \label{fig:hccm}
\end{figure}

Fig.~\ref{fig:hccm} partially visualises the hierarchical representation organisation analysed by SLINK as an icicle dendrogram~\cite{mcinnes2017hdbscan}, showing only hierarchical representation clusters containing more than 2,000 representations. Each displayed cluster is annotated with its cluster ID. The HCCM interpretations of these clusters are reported in Table~\ref{tab:hccm_results}. For a cluster matched with multiple predefined semantic classes, only the matched class with the highest L-score is reported.









\begin{table}[t]
\caption{HCCM interpretations of the hierarchical representation clusters visualised in Fig.~\ref{fig:hccm}.}
\vspace{2pt}
\label{tab:hccm_results}
\centering
\scriptsize
\setlength{\tabcolsep}{1.5pt}
\renewcommand{\arraystretch}{0.9}
\begin{tabular*}{\columnwidth}{@{\extracolsep{\fill}}clc@{\hspace{3pt}}clc@{}}
\hline
Cluster & Top1 Class & L-score &
Cluster & Top1 Class & L-score \\
\hline
47150 & -- & -- &
47158 & -- & -- \\

47151 & Female & 0.9993 &
47159 & Erik Estrada & 0.6029 \\

47152 & Male & 0.9994 &
47160 & Ernest Borgnine & 0.4462 \\

47153 & USA \& Female & 0.8708 &
47161 & Esai Morales & 0.3002 \\

47154 & Norway \& Female & 0.3779 &
47162 & UK \& Male & 0.5900 \\

47155 & USA \& Male & 0.6640 &
47163 & Eli Roth & 0.5648 \\

47156 & India \& Male & 0.9962 &
47164 & Eddie Izzard & 0.9704 \\

47157 & Ireland \& Male & 0.0916 &
& & \\
\hline
\end{tabular*}
\end{table}

Observing Fig.~\ref{fig:hccm} and Table~\ref{tab:hccm_results}, there are several cluster--class pairs achieve high L-scores. Specifically, clusters 47151 and 47152 are matched with the individual classes \emph{female} and \emph{male}, achieving near-perfect L-scores of 0.9993 and 0.9994, respectively. Clusters 47156 and 47164 are matched with the conjunctive class \emph{India \& male} and the identity-related individual class \emph{Eddie Izzard}, achieving high L-scores of 0.9962 and 0.9704, respectively. Moreover, none of the representation clusters in Fig.~\ref{fig:hccm} is interpreted using a nationality-related individual class when only the matched class with the highest L-score is reported, whereas six clusters are interpreted using nationality\&gender conjunctive classes instead. Lastly, those cluster--class pairs with low L-scores can be diagnosed. For instance, cluster 47154, interpreted as \emph{Norway \& female}, has an L-score of 0.3779. Further inspection shows that 0.3779 is limited by precision, as 61.21\% representations in this cluster do not belong to \emph{Norway \& female} but instead belong to other predefined semantic classes, especially \emph{UK \& female}.

\section{Conclusion}
\label{sec:conclusion}

This XAI work studies how the examined speaker recognition network hierarchically organises learned representations. SLINK is used to analyse the learned representations, yielding hierarchical representation clusters that collectively constitute the hierarchical representation organisation; CCM is used to evaluate the resulting clusters; and HCCM is used to interpret these clusters by finding their one-to-one matches with individual and conjunctive classes. The L-score quantifies the matching degree of cluster--class pairs in a way that makes mismatches diagnosable.

In CCM's evaluation, the SLINK-analysed representation organisation overall aligns strongly with individual classes related to speaker identity, gender, and nationality, even though the examined network is trained to recognise speaker identities rather than gender or nationality. Moreover, nationality\&gender conjunctive classes constructed in HCCM are more suitable for evaluating this representation organisation than nationality-related individual classes.

In HCCM's interpretation, gender-related individual classes match clusters at higher-level positions in the SLINK-analysed representation organisation, whereas identity-related individual classes and nationality\&gender conjunctive classes match clusters at remaining positions. Such interpretations provide insight into utterance semantics through the hierarchical representation organisation learned by the speaker recognition network and revealed by SLINK.

\section{Acknowledgement}
Mark D. Plumbley was supported by the Engineering and Physical Sciences Research Council (EPSRC) [grant number EP/Y028805/1]. For the purpose of open access, the authors have applied a Creative Commons Attribution (CC BY) licence to any Author Accepted Manuscript version arising.




\bibliographystyle{IEEEbib}
\bibliography{refs}

\end{document}